\documentclass[aps,nofootinbib,floatfix,onecolumn,11pt]{revtex4}

\usepackage{amsfonts,amsmath,amssymb}
\usepackage{graphicx,epsfig}

\def	\be	{\begin{equation}}
\def	\ee	{\end{equation}}

\def	\lf	{\left (}
\def	\rt	{\right )}

\begin{document}

\title{Enhanced Instability of de Sitter Space in Einstein-Gauss-Bonnet Gravity}

\author{Maulik Parikh}

\affiliation{Inter-University Centre for Astronomy and Astrophysics, 
Post Bag 4, Pune 411007, India} 

\begin{abstract}
\begin{center}
{\bf Abstract}
\end{center}
\noindent
We show that the addition of a topological Gauss-Bonnet term to the gravitational action can greatly increase the instability of four-dimensional de Sitter space, by favoring the nucleation of black holes. The pair-production rate given by the Euclidean action for the instanton takes the form $\exp(\Delta S)$ where $S$ is the entropy in Einstein-Gauss-Bonnet theory. The coefficient of the Gauss-Bonnet term in the action sets a stability bound on the curvature of empty de Sitter space. For that coefficient in the low-energy effective action of heterotic string theory, the maximal curvature of de Sitter space is in general much lower than the Planck scale.
\end{abstract}

\maketitle

\section{Introduction}

During the inflationary epoch, spacetime was well-described by de Sitter space. De Sitter space is known to be stable under classical perturbations \cite{ginspargperry}. Quantum-mechanically, however, de Sitter space has a nonperturbative instability due to the spontaneous nucleation of black holes \cite{ginspargperry,boussohawking}. But in Einstein gravity, the rate per unit volume of producing a black hole pair is only $e^{-\pi/\Lambda G}$, which is exceedingly low except when the effective cosmological constant, $\Lambda$, is close to the Planck scale. Moreover, any nucleated black holes have Planckian masses and hence evaporate in a flash.

In this paper, we show that adding a topological Gauss-Bonnet term to the gravitational action can greatly affect the pair-production rate of black holes; this is because the Euclidean black hole solution has a different global topology from Euclidean de Sitter space. We find that the Gauss-Bonnet term always enhances the non-perturbative instability of four-dimensional de Sitter space. The enhancement is by a factor of $e^{4 \pi \alpha/G}$ and, depending on the dimensionful coefficient, $\alpha$, of the Gauss-Bonnet action, can be enormous. The Gauss-Bonnet action appears at order $\alpha^\prime$ in the ten-dimensional low-energy effective action of heterotic string theory. Although the precise value of  $\alpha/G$ depends on the string coupling and the compactification down to four dimensions, given the likely bounds on the string scale the enhancement is at least of the order of $e^{(10^3)}$. Furthermore, the black holes thus formed are produced at the scale $\alpha$, rather than at the Planck scale, and are therefore longer-lived. Thus the presence of a Gauss-Bonnet term in the action has potential repercussions for Planck-scale inflation as well as for the production of primordial black holes.

\section{Black Hole Production in Einstein-Gauss-Bonnet Gravity}

In the path integral approach to quantum gravity, one considers a sum over gauge-inequivalent metrics:
\be
Z = \int {\cal D} g_{ab} e^{iI[g]  + {\rm gauge-fixing}} \; .
\ee
In ordinary quantum field theory, one Wick-rotates time to tame the divergence of the path integral. In general relativity, there is no preferred time coordinate but one can replace the sum over Lorentzian metrics with a sum over all ``Euclidean" metrics (more correctly Riemannian, i.e. positive-definite, metrics) that satisfy the appropriate boundary conditions. The resulting path integral still has dubious convergence properties because the gravitational action is indefinite and bounded neither from above nor from below. Nevertheless, this is the starting point of the Euclidean quantum gravity approach \cite{EQG}, a very successful program which is particularly well-adapted to nonperturbative semi-classical problems in quantum gravity. As in Yang-Mills theory, it is important that the sum includes solutions with different global topologies.

It is usually only feasible to evaluate the path integral in a saddle-point approximation. Then the path integral is given by a sum over gravitational instantons. These are defined as nonsingular geodesically complete solutions of the Euclidean gravitational equations, with or without a cosmological constant. The Euclidean action, $I_E$, for different gravitational instantons determines their relative probabilities. Ignoring the prefactor, we have
\be
\Gamma \sim \frac{\exp \lf -I_E[{\rm instanton}] \rt}{\exp (- I_E[{\rm background}])} \; .
\ee
This is interpreted as the rate per unit volume of spontaneously nucleating the object described by the instanton in the background spacetime. For example, the Euclideanized Schwarzschild solution and hot Minkowski space at the same temperature share the same boundary conditions. Then the action for the Euclidean Schwarzschild solution determines the rate per unit volume of spontaneously nucleating a black hole in thermal Minkowski space \cite{GrossPerry}.

The rate depends on the action. Now, since gravity is nonrenormalizable, one expects to find an infinite number of higher-curvature counter-terms in the action. Viewed as an effective action, the higher-curvature terms will generically play only a marginal role when the curvature of spacetime is small compared to the Planck scale. Nevertheless, as we shall see, higher-derivative terms can have an important effect on black hole production in de Sitter space. Here we will go to the next order in curvature, choosing the action to take the specific form
\be
I = \frac{1}{16 \pi G} \int d^D x \sqrt{-g} \lf R - 2 \Lambda + {\alpha} \lf R_{abcd} R^{abcd} - 4 R_{ab} R^{ab} + R^2 \rt \rt + I_{\rm boundary} \; .
\ee
This is the Einstein-Gauss-Bonnet action. The boundary terms, which generalize the Gibbons-Hawking term, vanish for the compact Euclidean manifolds we consider here so we will ignore them henceforth. The coefficient $\alpha$ is a constant of dimension $({\rm length})^2$. It is essential that $\alpha$ be positive; otherwise, the theory would admit black hole solutions with negative entropy \cite{myerssimon}. At quadratic order in curvature, there are three independent curvature invariants: $R^{abcd} R_{abcd}$, $R_{ab}R^{ab}$ and $R^2$; the motivation for picking the Gauss-Bonnet combination is two-fold: it is the only combination that is ghost-free (in all dimensions) when expanded around flat space, and it is precisely this combination that appears at order $\alpha^\prime$ in the low-energy effective action of heterotic string theory \cite{GrossWitten,Zwiebach}.

In four dimensions, the variation of the Gauss-Bonnet part of the action is a total derivative, canceling identically against the variation of the corresponding boundary term. These two terms together are in fact proportional to a topological invariant, the Euler character, $\chi$ \cite{Lanczos}. For compact four-dimensional manifolds, there are no boundary terms and so we have
\be
\chi  = \frac{1}{32 \pi^2}  \int d^4 x \sqrt{-g} \lf R_{abcd} R^{abcd} - 4 R_{ab} R^{ab} + R^2 \rt \; . \label{chi}
\ee
For Einstein spaces ($R_{ab} = \Lambda g_{ab}$), the Einstein-Gauss-Bonnet action for four-dimensional Euclidean compact manifolds is therefore
\be
I_E = - \frac{\Lambda V_4}{8 \pi G} - \frac{2 \pi \alpha}{G} \chi \; ,
\ee
where $V_4$ is the four-volume of the manifold.
 
The addition of a topological Gauss-Bonnet term obviously does not affect Einstein's equations in four dimensions, meaning that gravitational instantons of the Einstein-Hilbert theory continue to be solutions of the Einstein-Gauss-Bonnet theory. In particular, for positive $\Lambda$, Euclidean de Sitter space --- a four-sphere with radius $L = \sqrt{3/\Lambda}$ --- remains a gravitational instanton. The Euler character of an $S^4$ is 2 (as (\ref{chi}) confirms), and hence
\be
I_E[{\rm dS}] =  - \frac{\pi L^2}{G} - \frac{4 \pi \alpha}{G} \; .
\ee
Another solution is Euclidean Schwarzschild-de Sitter space whose line element is
\be
ds^2 = + \lf 1 - \frac{2GM}{r} - \frac{r^2}{L^2} \rt dt_E^2 + \lf 1 - \frac{2GM}{r} - \frac{r^2}{L^2} \rt^{\! \! -1} dr^2 + r^2 d\Omega^2 \; . \label{SdS}
\ee
This space has two potential conical singularities with radii $r_{\pm}$, which are found by solving a cubic. In Lorentzian signature, these correspond to the black hole and cosmological event horizon whose inverse Hawking temperatures are, respectively,
\be
\beta_{\rm hole} =  \frac{4 \pi r_- L^2}{(r_+ - r_-) (r_+ + 2r_-)} \; , \quad 
\beta_{\rm cosmo} = \frac{4 \pi r_+ L^2}{(r_+ - r_-) (2 r_+ + r_-)} \; . \label{beta}
\ee
When $r_+ \neq r_-$, the black hole and cosmological horizons have different temperatures, with $T_{\rm hole} > T_{\rm cosmo}$. Hence only one of the conical singularities can be eliminated by letting $t_E$ have the appropriate periodicity. Because of the remaining singularity, Euclidean Schwarzschild-de Sitter space is not in general nonsingular when $r_+ \neq r_-$ and is therefore not a gravitational instanton. 

However, in the extremal, or ``Nariai," limit the two horizons coincide: $r_{\pm} \to L/\sqrt{3}$. This happens for $M_{\rm Nariai} = L/\sqrt{27}$. By (\ref{beta}), both horizons then have zero temperature. At first sight, the region $r_- < r < r_+$ covered by the coordinates in (\ref{SdS}) seems to vanish in the extremal limit $r_- \rightarrow r_+$. However, this is an artifact of the coordinate system \cite{ginspargperry,boussohawking,extremalRN}. In fact, the four-volume for the Euclidean Schwarzschild-de Sitter black hole approaches a nonzero constant in the Nariai limit:
\be
V_4  = 4 \pi \beta_H \int_{r_-}^{r_+} r^2 dr = \frac{(4 \pi)^2 r_+ L^2 (r_+^3 - r_-^3)}{3(r_+ - r_-) (2r_+ + r_-)} \rightarrow \frac{16 \pi^2 L^4}{9}
\ee
Indeed, the Euclidean Nariai solution is a gravitational instanton, a completely regular space which has many interesting properties \cite{adventures}. As the largest black holes in de Sitter space, Nariai black holes also play a key role in resolving the de Sitter information puzzle \cite{dSinfo}. 
Pair-production of Nariai black holes in de Sitter space has been investigated before \cite{ginspargperry,boussohawking} in the context of Einstein gravity. There it was shown that a Euclidean Nariai black hole is geometrically and topologically the product of two two-spheres, $S^2 \times S^2$  \cite{ginspargperry,boussohawking}. Now, the Euler character for a product manifold (${\cal M} = {\cal M}_1 \times {\cal M}_2$) is the product of the Euler characters; this follows from the definition of the Euler character as an alternating sum of Betti numbers, $\chi = \sum_{n = 0}^D (-1)^n b_n$, and the K\"unneth formula for the Betti numbers of product topologies, $b_n({\cal M}) = \sum_{p + q = n} b_p({\cal M}_1) b_q({\cal M}_2)$.
Hence 
\be
\chi_{\rm Nariai} = 4 \; .
\ee
This can also be checked by inserting $R_{ab} = \Lambda g_{ab}$ and the Kretschmann scalar $R_{abcd}R^{abcd} = \frac{48 (GM)^2}{r^6} + \frac{24}{L^4}$ for Schwarzschild-de Sitter space directly into (\ref{chi}) and then taking the Nariai limit of coincident horizons. Thus we find that
\be
I_E[{\rm Nariai}] =  - \frac{2\pi L^2}{3G} - \frac{8 \pi \alpha}{G} \; .
\ee
The rate per unit volume of black hole pair production in the de Sitter background is therefore
\be
\Gamma \sim \frac{\exp \lf -I_E[{\rm Nariai}] \rt}{\exp (- I_E[{\rm dS}])}= \exp \lf - \frac{ \pi L^2}{3G}  + \frac{4 \pi \alpha}{G} \rt \; . \label{probability}
\ee
Since $\alpha > 0$, we see that the production rate of black holes is always enhanced in Einstein-Gauss-Bonnet gravity compared to the corresponding rate in Einstein gravity. Moreover, the amount of enhancement is highly sensitive to the value of $\alpha/G$.

\section{Discussion}

There is another way to understand (\ref{probability}). Without the Gauss-Bonnet term, the probability for black hole production can also be written as the exponent of the difference in Bekenstein-Hawking entropy between pure de Sitter space (with radius $L$) and extremal Schwarzschild-de Sitter space (which has two horizons with radius $L/\sqrt{3}$):
\be
\Gamma \sim \exp(-\pi L^2/3G) = \exp(\Delta A/4G)
\ee
Now when the Gauss-Bonnet term is included, the rate is no longer the difference in area. However, the entropy is proportional to the area only for black holes in Einstein gravity; for more general theories of gravity, one must use the Wald entropy \cite{wald}. In particular, for black holes in $D$-dimensional Einstein-Gauss-Bonnet gravity, the entropy is
\be
S = \frac{1}{4G} \int d^{D-2}x  \sqrt{\sigma} \lf 1 + 2 \alpha R_{(D-2)} \rt \; , \label{entropy}
\ee
where $R_{(D-2)}$ is the $D-2$-dimensional Ricci scalar on a horizon cross-section \cite{JacobMyers,myerssimon,stefano}. The first term is just the usual Bekenstein-Hawking area formula. The Gauss-Bonnet contribution to the entropy is given by the second term. In four dimensions, this is $\frac{\alpha}{2G} \int d^2 x \sqrt{\sigma} R_{(2)}$. But the two-dimensional Euler character is $\chi = \frac{1}{4 \pi} \int d^2 x \sqrt{\sigma} R_{(2)}$, which is $2$ for horizons whose spatial sections have spherical topology. Hence the Gauss-Bonnet contribution to the entropy is $4 \pi \alpha/G$, which exponentiates to precisely the modification to the nucleation rate. Thus, we find that, even in Einstein-Gauss-Bonnet gravity, the rate still takes the form
\be
\Gamma \sim \exp(\Delta S) \; ,
\ee
where $S$ is now given by (\ref{entropy}). It is interesting that the rate for black hole nucleation takes the same form as the emission rate for Hawking radiation from the de Sitter horizon, when backreaction effects are taken into account \cite{newcoords}; that the rate scales as the exponential of the entropy is consistent also with the idea that black hole entropy counts the number of quantum states \cite{garfinkle,secret}.

We see that, depending on the dimensionful scale $\alpha$, the presence of the Gauss-Bonnet term can dramatically enhance the production of black holes. Without the Gauss-Bonnet term, the probability for black hole production goes as $\exp(-\pi L^2/3G)$. The rate of black holes is therefore tremendously suppressed unless $L$ is of the order of the Planck length. But for Planckian curvatures, we would anyway be deep in the quantum gravity regime and outside the validity of a semi-classical analysis. However, with the Gauss-Bonnet term included, the production rate approaches unity not near the Planck scale but when $L^2 = 12 \alpha$. Thus if $\sqrt{\alpha} \gg l_P$, the nucleation rate could approach unity at a curvature scale that is significantly lower than the Planck scale.

One can estimate the value of $\alpha/G$ in string theory. The bosonic part of the ten-dimensional low-energy effective action of heterotic string theory is
\be
I = \frac{1}{2 \kappa_{10}^2} \int \! \! d^{10} x \sqrt{-g_s} e^{-2\phi} \left [R_s + 4 (\nabla \phi)^2 + \alpha^\prime \lf \frac{1}{8} {\cal L}_{\rm G-B} - 2 G^{ab} \nabla_a \phi \nabla_b \phi + 2 \Box \phi (\nabla \phi)^2 - 2 (\nabla \phi)^4 \rt \right ] 
\ee
where $\phi$ is the dilaton and the subscript $s$ indicates that the quantities refer to the string-frame metric \cite{Meissner}. (We have neglected terms associated with the Kalb-Ramond field, $B_{\mu \nu}$, and the gauge field strength, $F_{\mu \nu}$.) The higher-dimensional prefactor of the Gauss-Bonnet term is $\alpha^\prime/8$. To relate this to our coefficient $\alpha$, we have to transform from string frame to Einstein frame and reduce the dimensions from ten to four. Setting $\phi$ to zero, compactifying over a volume $V_6 = \mu (\alpha^\prime)^3$, and using $16 \pi G_{10} = (2 \pi)^7 g^2 {\alpha^\prime}^4$, we find
\be
\frac{\alpha}{G} = \frac{\alpha^\prime}{8G} = \frac{\mu}{64 \pi^6} \frac{1}{g^2} \; ,
\ee
where $g$ is the string coupling. For toroidal compactification at the smallest radius, the self-dual radius, we have $\mu = (2 \pi)^6$ and hence $\alpha/G = g^{-2}$; for larger compactification volumes, $\alpha/G$ is still larger. Indeed, if one takes $\sqrt{\alpha}$ to be around the string length then, since the string length is generically larger than the Planck length by at least an order of magnitude, $\sqrt{\alpha/G} \geq 10$, one finds that the enhancement factor is at least $e^{(10^3)}$.

The enhancement of black hole production in de Sitter space has consequences for inflationary cosmology. Inflation presumably cannot happen when the rate of black hole production is of order unity; black hole production thus places an upper bound on the effective cosmological constant:
\be
H_{\rm max} = \frac{1}{\sqrt{12 \alpha}} \; . \label{Hmax}
\ee
(Alternatively, one can regard this as a (weak) lower bound on the string coupling.) Now, although Nariai black holes have zero temperature, a small perturbation can separate their degenerate horizons. Then, since $T_{\rm hole} > T_{\rm cosmo}$, the black hole will eventually disappear leaving just de Sitter space. This makes sense entropically: empty de Sitter space has greater entropy than de Sitter space with a black hole in it. However, for an effective cosmological constant close to $1/\alpha$, the time between nucleation of black holes could exceed the time it takes for them to decay via Hawking radiation. There would then always be at least some black holes around in any given Hubble volume. This would render invalid the usual inflationary assumption of a spatially homogeneous background. Thus black hole production puts an upper bound on how curved empty de Sitter space can be; if it's too curved, it won't stay empty. 

Moreover, as these black holes are produced at the scale (\ref{Hmax}), they are larger than Planck-sized black holes roughly by a factor  of $\alpha/G$ and would correspondingly live longer by a factor of $(\alpha/G)^3$. Such black holes would therefore not only be produced later but would also live longer than Planck-sized primordial black holes; it would be interesting to study the time evolution of these black holes (and their electrically or magnetically charged cousins) to see whether any inflationary scenarios can be constrained \cite{primordial,boussohawking}.

It would also be interesting to see what effect adding a Gauss-Bonnet term has on the probabilities for other gravitational instantons \cite{GibbonsHawkinginst}. In general, the Gauss-Bonnet action affects the probability whenever the instanton has a different Euler number than the background. This could happen not only when the instanton is topologically nontrivial, but also when the background has nontrivial topology, as for example in elliptic de Sitter space \cite{gibbonselliptic,sanchez,edS}.

\vspace{0.3cm}
\noindent
{\bf Acknowledgments}

\noindent
We thank Saswat Sarangi, Jan Pieter van der Schaar, and Erick Weinberg.

\end{document}